\documentclass[aps,pre,twocolumn,showpacs,groupedaddress]{revtex4}

\usepackage[dvips]{graphics}

\begin{document}



\title{Lattice study of a Janus interface}

\author{Thomas A. McCormick}
\affiliation{Department of Chemistry, University of California at 
        Berkeley, Berkeley, CA 94720}
\date{\today}


\begin{abstract}
A lattice gas simulation of water between a hydrophobic plate and a
hydrophilic plate (a Janus interface)
shows large fluctuations in the number of liquid cells
in contact with the hydrophobic plate,
and a power spectrum similar
to the experimental results that Zhang, Zhu, and Granick found
[X. Y. Zhang, Y. X. Zhu, and S. Granick, Science \textbf{295}, 663 (2002)]
when measuring viscous response
in a Janus system.  Study of the spatial Fourier modes
of the liquid-vapor interface 
suggests that interface fluctuations with length scales between 
approximately 1.5 and 20 nm cause the effects observed in the simulation.
\end{abstract}


\maketitle

\section{\label{intro}Introduction}

The interactions between water and hydrophobic and hydrophilic surfaces
are important in many areas of research, including such diverse fields
as lubrication and protein folding.  
In order to study hydrophobic effects with as few additional complications
as possible, the behavior of water confined 
between two hydrophobic plates has been
the subject of many experimental and theoretical 
studies\cite{backg1, backg2, backg3, backg4, backg5, backg6, backg7}.  
More recently, experimentalists and theorists
have also examined the behavior of water confined between one hydrophobic
and one hydrophilic plate\cite{janus_gran, janus_bind}, a so-called
Janus interface, to learn how water reacts to the competing effects of
the plates.

While performing experiments to study
the response of water in a Janus system to
shear deformations, Zhang \textsl{et.al.}\cite{janus_gran}
found that the fluctuations
about the mean value of the viscous response were typically
25-50\% of the mean value, but ranged up to 100\%.  These
fluctuations are extraordinarily large
compared to the fluctuations observed when both plates are 
hydrophilic, and it is surprising that they do not average out
over the plates, which are approximately 10 $\mu$m on a side.
When the time series of the viscous response was Fourier analyzed, 
it was found that the resulting power 
spectrum is independent of frequency at low frequencies, but at
approximately $f = .001$ Hz, the power spectrum crosses over to power-law
dependence on frequency, with an apparent exponent near 
$f^{-2}$.  For frequencies greater than
$f = .01$ Hz, the power spectrum again levels off and
becomes seemingly independent of frequency.

It has been predicted\cite{lcw} that when water encounters a large
hydrophobic solute, such as the hydrophobic plate in a Janus system,
the water forms a semi-free liquid-vapor interface near the hydrophobic
surface.  The surface is unable to form hydrogen bonds with the water,
and the formation of an interface allows the water to form a more
energetically favorable hydrogen bonding network.  Zhang \textsl{et.al.}
suggested that their results are caused by fluctuations of 
such an interface.  They further identified
the time scale of $\sim 10^3/2\pi$ seconds, the inverse of the frequency
at which power-law behavior begins in their power spectrum, as the
lifetime of vapor or a vapor bubble.

This paper presents the results of a simulation designed to test the
hypothesis that fluctuations in the liquid-vapor interface are the 
source of the large fluctuations and power spectrum seen experimentally.
We chose to study a lattice gas because it is
a simple model system that can be used to examine the 
large length scale behavior of a liquid 
between competing solvophobic and solvophilic plates, and can also
describe interface fluctuations.  
After describing the details of the simulation 
in Section \ref{sim},
results are presented in Section \ref{results}, and the paper is
concluded with a brief discussion.


\section{\label{sim} Simulation Details}

\begin{figure}
\includegraphics{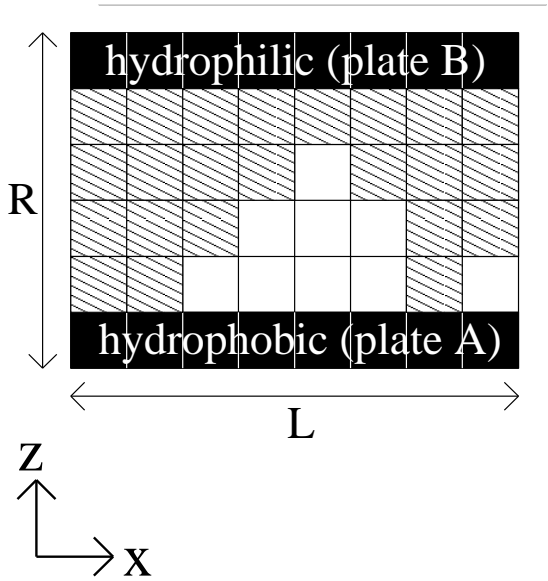}
\caption{\label{lattice_setup_fig} Cross section of the lattice in
the simulation.  A possible configuration of vapor (white) cells and 
liquid (shaded) cells is also shown.}
\end{figure}

The system in the simulation is an $L\times L \times R$ cubic lattice
gas, where $L$ is the side length of the hydrophobic and
hydrophilic plates in the $x$ and $y$ directions in units of the
lattice spacing $a$, and
$R$ is the distance between the plates, in the $z$ direction, as
shown in Figure \ref{lattice_setup_fig}.  $L$ and $R$ were chosen to
be 128 and 8, respectively.  Each cell $i$ has an associated 
variable $n_i$ which labels the cell as liquid ($n_i=1$)
or vapor ($n_i=0$).  The cells around the edges of the system 
are constrained to have $n_i=1$
in order to ensure that there is no evaporation between the plates.
The cells in the $z=0$ and $z=7$ layers represent the hydrophobic
and hydrophilic plates, respectively, and are constrained to have
$n_i=1$ at all times so that liquid in the bulk always interacts
with the plates.
The remaining unconstrained cells in the bulk obey the Hamiltonian 

\begin{equation}
\label{ham}
\beta H = -\beta\epsilon \!\! \sum_{<ij>} \!\! n_i n_j \
          -\beta\mu \sum_i n_i \
          -\beta\sigma_A \!\!\!\!\!\! \sum_{i \:\: n.n.\, A} \!\!\!\! n_i \
          -\beta\sigma_B \!\!\!\!\!\! \sum_{i \:\: n.n.\, B} \!\!\!\! n_i
\end{equation}

\noindent where $\beta = (k_BT)^{-1}$, $k_B$ is Boltzmann's constant,
$<ij>$ indicates a sum over
nearest neighbors, and $i\:\: n.n.\, A$ and $i\:\: n.n.\, B$ 
indicate sums over cells adjacent to plate A (the hydrophobic plate)
or B (the hydrophilic plate).

The parameters $\beta\epsilon$ and $\beta\mu$ have the same
values as in the simulation of Luzar and Leung\cite{backg7}, 
$\beta\epsilon = 1.26$ and $\beta\mu = -6\epsilon/2 + 1.84 \times 10^{-4}$, 
chosen to match the surface tension and isothermal compressibility of
water at 300 K.  With these parameters, the lattice spacing $a$
corresponds to .193 nm, so the plates are approximately
1.5 nm apart, within the range studied by Zhang \textsl{et.al.}
However, the plates in the simulation have a side length of
approximately 25 nm, which is significantly
smaller than the experimental plates.
The interactions with the plates are $\beta\sigma_B = \beta\epsilon$ and
$\beta\sigma_A = .1\beta\epsilon$, which gives a contact angle of 
143$\,\mathring{ }$ at the hydrophobic plate, comparable to the
contact angle of approximately 120$\,\mathring{}$ in the experimental
system. 

Zhang \textsl{et.al.} were careful to show that although they
applied shear forces to the hydrophobic plate
and measured the response of the water in their experiments, 
the system was still in the
linear response regime, so fluctuations present in the
nonequilibrium 
experiment should be present at equilibrium as well.  
To make the simulation as simple as possible, we chose to
study an equilibrium lattice gas propagated by Metropolis Monte Carlo,
according to the Hamiltonian in equation (\ref{ham}) for 
approximately 140,000 passes through the lattice.  
The contact density, defined to be the
number of liquid cells in contact with the hydrophobic plate, 

\begin{equation}
\label{def_coft_eqn}
c(t) = \sum_{i \:\: n.n. \, A} \!\!\!\! n_i(t)
\end{equation}

\noindent was
calculated as a function of time, where time is measured in units
of Monte Carlo passes through the lattice.
$c(t)$ should be related to the viscosity of water near
the hydrophobic plate, because the viscosity measured at the plate
should increase with the amount of liquid (rather than vapor) in
contact with the plate.

The distance of the liquid-vapor interface $h$ from the
hydrophobic plate, as a function of 
$(x,y)$ and time, was also calculated in the simulation.  
The interface in the $(x,y)$ 
column is defined to be located at the average of the distances of
the liquid cell closest to, and the vapor cell
farthest from, the hydrophobic plate in that column, reminiscent
of Weeks' derivation of the capillary wave 
model\cite{cap_wave, def_interf}:

\begin{equation}
h(x,y) = \frac{1}{2} \left[ \max_{(x,y)} \left[ z_i(1-n_i) \right] \
                     + \min_{(x,y)} \left[ z_i n_i \right] \right]
\end{equation}

\noindent
The value of $h$ when liquid is in contact with the hydrophobic
plate is $h = (0+1)/2 = 0.5$, and increases in increments of $1$
to its maximal value of $h = (6+7)/2 = 6.5$,  
for vapor in contact with the hydrophilic plate.  

The fact that the edges 
of the system are constrained to be liquid for all $z$
leads to a boundary region of approximately 12 cells from each edge
where the fluctuations of $h$
are smaller than those in the middle of the lattice.
Results reported below for the $104\times104\times8$ lattice 
resulting from ignoring the outer perimeter of 12 cells are
virtually indistinguishable from results with a boundary of
32 cells, so the lattice seems to be 
large enough to study a Janus interface.

Inside the boundary region, the time and space-averaged interface 
position is $\langle h \rangle = 2.0$, which means that on average,
the first one or two layers in contact with the hydrophobic plate
are vapor.  Because the plate is so significantly dewetted, changes
in the amount of liquid in contact with the plate in our simulation
are caused largely by fluctuations of the liquid-vapor interface, 
rather than by spontaneously-forming vapor bubbles.  
To the extent that this lattice gas
is an accurate description of the experimental Janus system, then,
spontaneously-forming vapor bubbles do not seem to be
important contributors to the results.


\section{\label{results} Results}

\begin{figure}
\includegraphics{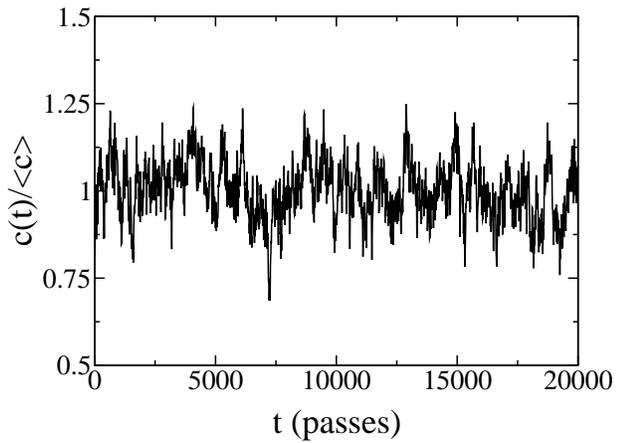}
\caption{\label{coft_fig} A portion of the time series of the contact
density, normalized by the time-averaged value 
$\langle c\rangle = 2673$.  The value of the
average contact density indicates that
approximately 25\% of cells in the $z=1$ layer have $n_i=1$ at any
given time.}
\end{figure}

\begin{figure}
\includegraphics{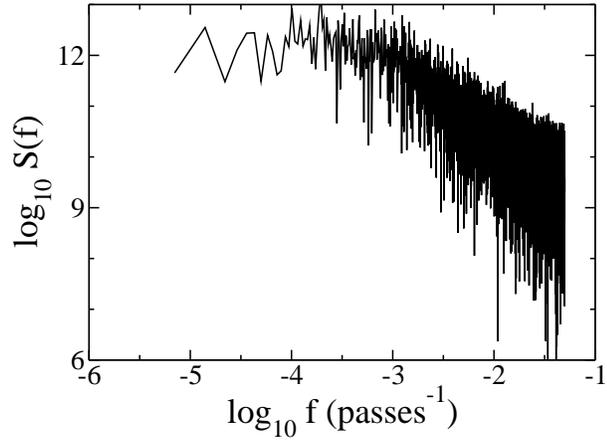}
\caption{\label{cpow_fig} The power spectrum $S(f)$ resulting from the
time Fourier transform of $c(t)$. }
\end{figure}

The contact density $c(t)$ and interface $h(x,y;t)$ were saved every
10 passes through the lattice during the simulation.
A section of the time series for the contact density measured over the
central $104\times104\times8$ lattice is shown in Figure \ref{coft_fig}.  
The fluctuations in $c(t)$ range up to approximately 25\% of
the mean, which is within the range seen experimentally.
The power spectrum resulting from the time Fourier transform of
$c(t)$, $S(f) = \left| \tilde{c}(f) \right|^2$, where

\begin{equation}
\tilde{c}(f) = (\Delta t)\sum_{t=0}^{t_{max}} c(t) \exp[2\pi ft/t_{max}]
\end{equation}

\noindent and $\Delta t$ is the time between samples (10 passes in 
our simulation) is shown on a log-log plot in Figure 
\ref{cpow_fig}.  The power is independent of frequency until approximately 
$f = .001 \;\mathrm{passes}^{-1}$, where it begins to drop off.  
The best-fit line to the data between 
$f = .001 \;\mathrm{passes}^{-1}$ and 
$f = .015 \;\mathrm{passes}^{-1}$ has a slope of $-1.48 \pm .04$, less
than the apparent power-law of $f^{-2}$ seen experimentally.  There
seems to be a slight leveling off of the power as a function of 
frequency at high $f$, but the data are so noisy that it is difficult
to quantify.  The
contact density $c(t)$ thus exhibits behavior surprisingly similar to
the viscous response in the experiments of Zhang \textsl{et.al.}

In order to try to understand the cause of the fluctuations and power
spectrum of $c(t)$, the dynamics of the liquid-vapor interface were
studied.  The characteristic length scales of fluctuations of the 
liquid-vapor
interface relevant to fluctuations in the contact density can be 
inferred from an analysis of the spatial Fourier modes of the 
interface.  The Fourier modes $\hat{h}(k_x,k_y;t)$ of the interface
can be defined with the equation

\begin{equation}
\label{space_ft}
\hat{h}(k_x,k_y;t) \equiv (\Delta x)^2 \sum_{x,y=0}^N
            h(x,y;t) e^{-2\pi i (k_x x + k_y y)}
\end{equation}

\noindent where $\Delta x$ is the spatial resolution ($a$ in the
simulation) and $N$ is the number of cells in each dimension (104
in the simulation).

\begin{figure}
\includegraphics{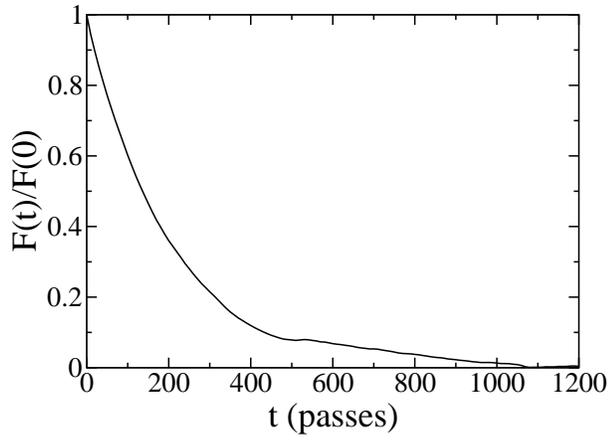}
\caption{\label{dFdF_fig} The time autocorrelation function for 
fluctuations in the $\mathbf{k}=(0,0)$ spatial Fourier mode of the
liquid-vapor interface.}
\end{figure}

Figure \ref{dFdF_fig} shows the time autocorrelation function of the
$\mathbf{k}=(0,0)$ spatial Fourier mode of the interface,

\begin{equation}
F(t) = \left\langle \delta \hat{h}(0,0;0) \
                    \delta \hat{h}(0,0;t) \right\rangle 
\end{equation}

\noindent where $\delta \hat{h}(k_x,k_y;t) = \hat{h}(k_x,k_y;t) - \
\langle \hat{h}(k_x,k_y;t)\rangle$,
as a function of time.  $F(t)$ decays on a time scale
of approximately 1000 passes, which is the inverse of the frequency
near which power-law behavior begins in $S(f)$.  In the sense that
fluctuations of the whole interface correspond to changes in the 
amount of vapor near the hydrophobic plate, 
$f \sim .001 \;\mathrm{passes}^{-1}$ corresponds to the lifetime
of vapor fluctuations, similar to the conclusion of Zhang 
\textsl{et.al.}  The fact that the relaxation time of fluctuations
of the entire interface corresponds to the frequency at which
power-law behavior begins in $S(f)$ suggests that interface 
fluctuations, rather than spontaneously-forming vapor bubbles,
are the cause of the fluctuations in $c(t)$.

If $D$ is the smallest length scale of fluctuations in
the interface relevant to fluctuations in the contact density, 
then contributions to $h(x,y;t)$ from Fourier modes with 
wavelengths smaller than $D$ should be minimal.  A new function,
$h_D(x,y;t)$, with no contributions from short-wavelength
Fourier modes, can be created by taking the inverse transform
of equation (\ref{space_ft}), but cutting off the sum over
wavevectors at $k_D = 1/D$:

\begin{equation}
h_D(x,y;t) = \frac{1}{N^2} 
           \sum_{k_x=0}^{k_D} \sum_{k_y=0}^{k_D} \
           \hat{h}(k_x,k_y;t) e^{2\pi i k_x x} e^{2\pi i k_y y}.
\end{equation}
When the small-wavelength Fourier modes are not included,
$h_D(x,y;t)$ is no longer restricted to take on discrete values.
The lack of high-frequency Fourier modes also smoothes out
the interface sufficiently that it seldom comes in contact with
the hydrophobic plate, so the definition of $c(t)$ in equation
(\ref{def_coft_eqn}) is no longer useful in the context of
fluctuations of $h_D(x,y;t)$.  Instead, a clipping
function is used to create a function $c_D(t)$,
\begin{equation}
\label{def_cD}
c_D(t) \equiv \sum_{x,y} \Theta[ \lambda - h_D(x,y;t) ]
\end{equation}
where $\Theta$ is the Heaviside function, and the smooth interface
$h_D(x,y;t)$ is considered to be in contact with the hydrophobic
plate when $h_D(x,y;t)\le\lambda$.  $\lambda$ is chosen for each
$D$ so that the time average of $c_D(t)$ equals the 
time average of the total contact density $c(t)$. 

To estimate the smallest-wavelength fluctuations of the
interface relevant to the fluctuations in the $c(t)$ time series
that was collected during the simulation,
the functions $h_D(t)$ with 
$D = 6,7,8,9,10,12,14, \mathrm{and} \: 16 \: a$ 
were computed from
the time series of $h(x,y;t)$ collected during the simulation, and
the corresponding $\lambda$'s ranged from $\lambda=1.50$ for $D=6 \: a$,  
to $\lambda=1.67$ for $D=16 \: a$.
The power spectra $S_D(f) = \left| \tilde{c}_D(f) \right|^2$ were 
computed, and the slope of each curve was measured on a log-log plot
between $f=0.001 \;\mathrm{passes}^{-1}$ 
and $f=0.015 \;\mathrm{passes}^{-1}$, to see how many Fourier modes
could be removed from $h(x,y;t)$ but still obtain a 
slope statistically
indistinguishable from the slope of $S(f)$ over the same frequency
range.  It should be noted that this 
calculation was carried out as a technique to analyze the data
collected during the simulation, and not to make a statement about
the exact power-law behavior of $S(f)$, since the data are
quite noisy.  
The value of the slope for each $D$ is displayed
in Table \ref{slope_table}.  The largest value of $D$ with a slope
still within one standard deviation of the true slope of $S(f)$,
$-1.48 \pm 0.04$, is $D=8 \: a$, which suggests that fluctuations
with wavelengths smaller than $8\: a$ are not important to the
fluctuations in $c(t)$.  

\begin{figure}
\includegraphics{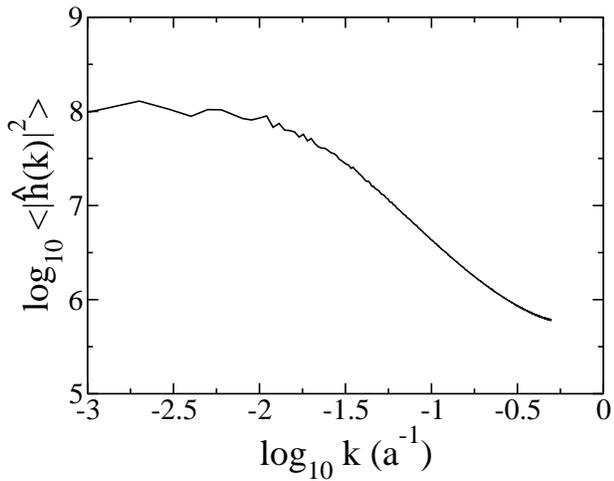}
\caption{\label{interf_pow_fig}Time averaged power spectrum of interface
fluctuations as a function of $k=|\mathbf{k}|$.
}
\end{figure}

When a similar calculation was performed where the small-$|\mathbf{k}|$
(large-wavelength) Fourier modes of the interface were progressively
removed from the sum in equation (\ref{space_ft}), the slope of the
resulting power spectrum was far outside of the standard deviation of
the true slope.  In order to estimate the largest length scale of
interface fluctuations relevant to fluctuations in $c(t)$, a short
simulation was run on a larger lattice.  A simulation identical to
that described in Section \ref{sim}, but with $L=1024$, was run
for approximately 2000 
Monte Carlo passes through the lattice.  The power 
spectrum of the interface fluctuations, $|\hat{h}(k_x,k_y;t)|^2$
was calculated every 10 passes, 
and the time-averaged power spectrum
is shown in Figure \ref{interf_pow_fig}, plotted versus $k=|\mathbf{k}|$.  
The power begins to depend on $k$ at approximately $k=.01 \: a^{-1}$,
which suggests that fluctuations over length scales up to approximately
$100 \: a$ are relevant in the simulation.  Thus, it seems 
that interface fluctuations
over length scales between $8\:a$ ($\sim$ 1.5 nm) and 
$100 \: a$ ($\sim$ 19 nm) cause the fluctuations in $c(t)$ in
the simulation.

Additional simulations were performed to determine the effects of
the parameters of the model upon the results described above.
Simulations at higher and lower temperatures (corresponding to
physical temperatures between 290 and 320 K), and with different 
values of $\beta\sigma_A$, were carried out.  $\beta\epsilon$ and
$\beta\mu$ were not varied independently so that the system would
remain close to liquid-vapor coexistence, and $\beta\sigma_B$ was
not varied because the interface is so far from the hydrophilic
plate that the results would be unchanged.

The results of these parameter changes were consistent with 
physical intuition.  As temperature increases, the interface
fluctuates more, causing larger-magnitude fluctuations in $c(t)$
and also causing $\langle h\rangle$ to increase (that is, the
interface moves farther from the hydrophobic plate).  The
power in the spatial Fourier modes of the interface increases,
and the leveling off of the power spectrum at small
$k$ moves to smaller $k$, indicating
that larger length-scale interface fluctuations are important, as
would be expected.  The behavior of $S(f)$ does not change noticeably
over the temperature range studied, and the linear regime begins
at approximately the same value of $f$.

$\beta\sigma_A$ was varied from 0.0 to $0.2\beta\epsilon$.
Increasing $\beta\sigma_A$ corresponds to making the hydrophobic
plate more hydrophilic, and as would be expected, doing so
decreases $\langle h\rangle$ and the magnitude of the fluctuations
in $c(t)$, and increases $\langle c\rangle$.  The power in the
interface Fourier modes decreases substantially, and the power
spectrum levels off at much larger values of $k$, indicating that 
large-wavelength Fourier modes become less important in the 
interface dynamics, which is reasonable since the interface is
so close to the plate and thus has little room in which to move.
When $\beta\sigma_A$ is increased to
$0.2\beta\epsilon$, $\langle h\rangle = 0.89$, indicating that
the hydrophobic plate is almost completely wetted, and 
the linear regime in $S(f)$ disappears.

\begin{table}
\caption{\label{slope_table} Slope of the linear regime on a log-log
plot of $S_D(f)$ for different values of $D$ (the units of D are
lattice spacings $a$).  The slopes have a standard deviation of 
approximately 0.04.}
\begin{ruledtabular}
\begin{tabular}{p{2.0cm}p{2.0cm}p{2.0cm}}
$D$ & slope\\[0.5ex]\hline
6 & -1.50 \\
7 & -1.48 \\
8 & -1.46 \\
9 & -1.42 \\
10 & -1.41 \\
12 & -1.40 \\
14 & -1.36 \\
16 & -1.34
\end{tabular}
\end{ruledtabular}
\end{table}


\section{\label{disc}Discussion}

The results of the simulation described above
are very similar to the experimental
behavior of water at a Janus interface, and support the hypothesis
that the large
fluctuations and power spectrum of viscous response that were seen
experimentally can be attributed to the fluctuating liquid-vapor
interface interacting with competing hydrophobic and hydrophilic walls.
The power spectrum of fluctuations in $c(t)$
has apparent power-law behavior at intermediate
frequencies, but the power law is a less rapidly-decaying function of
frequency then the apparent experimental 
value of $f^{-2}$.  However, the combination of the noise in the power
spectrum from the simulations, and the lack of a precise experimental
value, means there is no firm basis for a detailed discussion
of the exact value of the exponent. 
Regardless of that exact value, a Fourier analysis of the liquid-vapor
interface suggests that fluctuations of the interface over length 
scales between 1.5 and 19 nm cause the fluctuations in $c(t)$ that 
were observed in the simulation.  It is probable that fluctuations
over these length scales would be averaged out 
in a simulation of a system with the much larger
plates studied experimentally, but it is interesting that such
large fluctuations could be observed in such a simple system.

\begin{acknowledgments}
A conversation with S. Granick motivated this work, and the author would
like to thank D. Chandler for many helpful discussions.
This
research has been supported by a grant from the Department of Energy
(DEFG03 99ER14987).
\end{acknowledgments}



\newpage

\vspace{2.0cm}

\vspace{2.0cm}

\newpage

\vspace{2.0cm}

\newpage

\vspace{2.0cm}
\newpage

\end{document}